\documentclass[prb,onecolumn,showpacs,preprintnumbers,amsmath,assume]{revtex4}
\usepackage{graphicx}
\usepackage{dcolumn}
\usepackage{bm}
\pdfoutput=1

\begin{document}

\title{Perfectly matched layers simulation in 2-D \textbf{VTI} media}

\author{P. Contreras, G. Larrazabal, and C. Florio}
\affiliation{Multidisciplinary Center for Scientific Visualization and Computation (CEMVICC)
Science Faculty Universidad de Carabobo, Carabobo, Venezuela.}
\date{\today}

\begin{abstract}
Implementing computational boundary conditions, such as
perfectly matched layers $\textbf{PML}$ does have advantages
for forwarding modeling of the earth's crust. The mathematical
modeling of many physical problems encountered in industrial
applications often leads to a system of linear partial
differential equations \textbf{PDEs}. It considerably improves
the visualization of seismic events relevant to oil and gas
exploration. In this work, we present an efficient numerical
scheme for hyperbolic partial differential equations where a
computational technique takes care of reflections at the
border´s domain using a linear \textbf{2-D} elastic-wave system
of decoupled equations with \textbf{PML}-type boundary
conditions. The key idea is to introduce a layer that absorbs
the reflections from the borders improving images
visualization. Anisotropy has been reported to occur in the
Earth's three main layers. The crust, the mantle, and the core;
but this implementation refers to the case of a vertical
transversal anisotropic medium \textbf{VTI} in the crust-layer.
Images and screen-shots of the longitudinal \textbf{P}
impulse-response and the \textbf{SV} transverse
impulse-response are obtained at different times. This
computational method enables us to achieve images for the
\textbf{P} and \textbf{SV} response-impulses, and to obtain
high quality synthetic seismograms for the \textbf{PP} and
\textbf{PS} reflection events in a 2-D \textbf{VTI} two-layer
model.

\end{abstract}

 \maketitle

\section*{\textbf{Introduction}}

The wave equation plays an important role in seismic oil and
gas exploration since it allows modeling the earth structure.
In the past decades, seismic anisotropy has been gaining
attention from academic and industry, in part thanks to
advances in anisotropy parameter estimation
\protect\cite{gre3,gre2}

A material is said to be anisotropic if the value of one or
more of its properties varies with direction, consequently
anisotropy consideration improves the subsurface imaging.
Seismic Anisotropy can be defined as the dependence of seismic
velocity on a direction or upon an angle. Anisotropy is
described by a $4^{th}$ order elasticity tensor with 21
independent components for the lowest-symmetry case
\protect\cite{hel,mus,lan,del}. However in practice,
observational studies are unable to distinguish all 21 elements
and anisotropy considerations are usually simplified. For
seismic exploration, the most complicated case occurs in
fractured monoclinic media, with 9 elastic constants
\protect\cite{gre}. In general, two or more sets of vertical
non-corrugated and not perpendicular fractures produce an
effective monoclinic medium with a horizontal symmetry plane.
The second most important application occurs for the
orthorhombic model. The orthorhombic model describes a layered
medium fracture in two orthogonal directions
\protect\cite{tsv,con2}, However in the simplest form that
seismic exploration uses it, there are two main kinds of
transverse anisotropy \textbf{TI}. One is called horizontal
transverse isotropy-\textbf{HTI} which is a common model in
shear-wave studies of fractured reservoirs that describes a
system of parallel vertical penny-shaped cracks embedded in an
isotropic host rock\protect\cite{con1}, henceforth this kind of
anisotropy is associated with cracks and fractures. The other
one is called vertical transversal anisotropy-\textbf{VTI}
\protect\cite{thom} and it is associated with layering and
shales. Sometimes people call it vertical polar anisotropy.

In the beginning, forward modeling was done by simulating the
scalar $P$ wave field obtained from the acoustic wave equation.
However, the earth-crust is elastic and anisotropic
\protect\cite{hel,gre,con2} and all propagation modes should be
considered in order to observe anisotropy effects. Forward
modeling and parameter estimation are almost the most
fundamental to all other anisotropy applications in oil
exploration \protect\cite{gre2}. In part, this can be done by
numerically solving the hyperbolic elastic wave equation
\textbf{WE}. The vector and tensor fields which represent the
\textbf{WE} improves the information about the 3D earth-crust
geology. This translates into a better subsurface imaging and
it is of extreme importance for the oil and gas seismic
exploration.

The most common type of anisotropy that occurs in the earth's
crust is the vertical transverse anisotropy \textbf{VTI} which
is observed in sedimentary rocks \protect\cite{winterstein}. To
achieve 3D seismic modeling with \textbf{VTI} anisotropy
requires the knowledge of five elastic constants. However in
two dimensions only four constants are involved, namely
$C_{11}$, $C_ {33}$, $C_ {13}$ and $C_ {44}$ (in this work we
use Voigt notation for the elastic constants)
\protect\cite{faria,thom}. Henceforth shear wave splitting is
not considered in a 2-D modeling because of the lack of the
elastic constant $C_{66}$. Abundant geological evidence of
shales shows the importance of \textbf{VTI} models in seismic
reservoir characterization \protect\cite{gre3,gre2}.

In this work, we implement an algorithm using a linear finite
difference decoupled \textbf{PDEs} in terms of the velocities
and stress tensor components. To implement such a model we use
central-difference second order approximation for space
variables and time partial derivatives. Henceforth we use the
\textbf{finite-difference time-domain} \textbf{FDTD} method
\protect\cite{yee}. In addition, we use staggered stencils for
computational storage \protect\cite{virieux}. Staggered cells
grant the calculation of different physical quantities at
different mesh-grid points (see Figure \ref{f1}), reducing the
computational cost storage. Additionally, with this algorithm
we can introduce different source-receiver configurations,
making it ideal for simulation in oil and gas seismic modeling.

To implement a \textbf{FDTD} solution of the \textbf{WE}, a
computational frame must first be established. The
computational domain is the physical region over which the
simulation is performed. To achieve the condition of
non-reflective borders, we use the artificial technique of
perfectly coupled layers \textbf{PML}
\protect\cite{festa,dimitri}. Consequently, the effect due to
the reflection of the waves at the borders of the domain is
automatically reduced.

\textbf{PML} implementation in seismic exploration requires a
reformulation of the linear \textbf{WE} to eliminate the
unwanted reflections. Thus, this work is structured in the
following way: Section \textbf{I} introduces the topic of
seismic anisotropy and its relevant application in R\&D Oil
industry. Section \textbf{II} briefly describes the
\textbf{PDEs} for a 2-D medium with \textbf{VTI} shale
anisotropy, and the theoretical implementation of the
\textbf{PML} obtaining a decoupled \textbf{EWs} (elastic
equation system) including the new border computational
conditions. Section \textbf{III} describes the numerical
implementation that will allow CPU time reduction.
Subsequently, we find the correct decoupled finite-difference
scheme in terms of the staggered-stencil mesh taking into
consideration the \textbf{PML} (see Figure \ref{f1}). Finally,
in section \textbf{IV} the outcome shows the applicability of
this technique. We were able to achieve very neat and sharp
subsurface images without reflection events at the edges of the
2-D earth-crust two dip layer shale model.

\section*{\textbf{2-D linear \textbf{VTI} media}}

As we stated before, the formulation of a linear \textbf{WEs}
in terms of temporal derivatives of the velocity and the stress
fields can be used to propagate waves. The \textbf{FDTD}
technique reproduces the fields forward in time-domain. This is
called forward modeling. This procedure is designed through a
3D staggered finite-difference grid \protect\cite{faria}. This
formulation is widely used in the literature for seismology
purposes \protect\cite{virieux}. In 2-D \textbf{VTI}
anisotropy, it involves the transformation of five coupled
first-order \textbf{PDEs}, namely the two general equations of
motion for the vector velocity $(v_x,v_z)$ field which are:
\begin{displaymath}
\rho \frac{\partial v_x}{\partial t} = f_x + \frac{\partial
\sigma_{xx}}{\partial x} + \frac{\partial \sigma_{xz}}{\partial
z},
\qquad
\rho \frac{\partial v_z}{\partial t} = f_z +
\frac{\partial \sigma_{xz}}{\partial x} + \frac{\partial
\sigma_{zz}}{\partial z},
\end{displaymath}
where $\rho$ is a constant density, $f_x$ and $f_z$ are
external forces driven by the source. Since we use the
\textbf{FDTD} time-domain technique, a seismic pulse is used as
the source, then the response of the system over a wide range
of frequencies can be obtained with a single simulation.

And the three equations for the stress $\sigma_{ij}$ second
order tensor field \vspace{-0.5cm}
\begin{center}
\[
   \sigma_{ij}=
  \left[ {\begin{array}{cc}
   \sigma_{xx} & \sigma_{xz} \\
   \sigma_{zx} & \sigma_{zz} \\
  \end{array} } \right]
\]
\end{center}
where only four elastic constants are used as the input
variables for the \textbf{WEs}:
\begin{displaymath}
\frac{\partial \sigma_{xx}}{\partial t} = c_{11} \;
\frac{\partial v_x}{\partial x} + c_{13} \; \frac{\partial
v_z}{\partial z}, \qquad \frac{\partial \sigma_{zz}}{\partial
t} = c_{33} \; \frac{\partial v_z}{\partial z} + c_{13} \;
\frac{\partial v_x}{\partial x}, \; \qquad \frac{\partial
\sigma_{xz}}{\partial t} = c_{44} \; \Big[\frac{\partial
v_x}{\partial z} + \frac{\partial v_z}{\partial x}\Big],
\end{displaymath}
the longitudinal $P$ and the transverse $SV$ modes given by the
previous five \textbf{PDEs} are coupled\protect\cite{faria}.

\subsection*{\textbf{2-D \textbf{VTI} decoupled \textbf{PDEs} with \textbf{PML}}}

In this subsection, the \textbf{PML} technique is briefly
explained and applied to the 2-D modeling. In order to make the
finite difference method more stable and convergent, we follow
the arguments presented previously\protect\cite{festa,dimitri}
for the stability and the dispersion conditions in a 2-D
\textbf{EWs} (we should mention that a 3D anisotropy
implementation has not been achieved with this technique, due
to failures in the stability and dispersion conditions when
considering anisotropy\protect\cite{faria}). Even in works
dedicated to seismology, the use of this technique helps to
improve 2-D imaging resolution at larger scales. The
seismological domains have borders as it happens in seismic
exploration. \textbf{PML} in this case, reduces the reflections
at the borders in a significant amount compared with the use of
traditional mathematical boundary conditions .

The \textbf{PML} mathematical implementation consist in the
introduction of a modified coordinate system, where the
expansion coefficient is a complex number with an evanescent
imaginary part. This generalization is achieved through the
following substitution \vspace{-1.3cm}
\begin{center}
\[
\frac{\partial}{\partial x} \rightarrow \frac{1}{1 + i \; \frac{p(x)}{\omega}} \; \frac{\partial}{\partial x},
\]
\end{center}
and \vspace{-1.2cm}
\begin{center}
\[
\frac{\partial}{\partial z} \rightarrow \frac{1}{1 + i \; \frac{p(z)}{\omega}} \; \frac{\partial}{\partial z},
\]
\end{center}
where $p(x)$ and $p(z)$ are the coefficients of the
\textbf{PML}. They are given by the following expressions
\protect\cite{dimitri}
\begin{displaymath}
 p(x) = p_0 \; (x/L)^N, \; and \; p(z) = p_0 \; (z/L)^N,
\end{displaymath}
with $p_0$ $=$ $-3 \; v_p \; log R_c/(2 \;L)$, where $L$ is the
thickness of the perfectly coupled layer, $N$ the size of the
problem and $p_0$ has an approximate value of 341.9

When the above coordinate system is replaced in the equations
for the \textbf{VTI} medium of the previous section, the system
becomes a new linear decoupled \textbf{PDE}-\textbf{EWs}.
Because of that replacement appears new equations for $v^1_x$
and $v^2_x$ new decoupled velocities:
\begin{displaymath}
\rho \; \Big[ \frac{\partial }{\partial t} + p(x) \Big]  v^1_x = f_x + \frac{\partial \sigma_{xx}}{\partial x},
\qquad
\rho \; \Big[ \frac{\partial }{\partial t} + p(z) \Big] v^2_x = \frac{\partial \sigma_{xz}}{\partial z},
\end{displaymath}
For the decoupled $v^1_z$ and $v^2_z$ velocities the equations
are:
\begin{displaymath}
\rho \; \Big[ \frac{\partial }{\partial t} + p(x)\Big] v^1_z =
f_z + \frac{\partial \sigma_{xz}}{\partial x}, \qquad \rho \;
\Big[ \frac{\partial }{\partial t} + p(z)\Big] v^2_z =
\frac{\partial \sigma_{zz}}{\partial z}.
\end{displaymath}
These decoupled velocities~\protect\cite{dimitri} are related
to the coupled ones by the relation $v_x = v^1_x + v^2_x$ and
$v_z = v^1_z + v^2_z$.

The stress tensor field is mathematically treated in the same
way, obtaining a decoupled system for the new stress equations:
The new decoupled $\sigma^1_{xx}$ and $\sigma^2_{xx}$ stress
equations are:
\begin{displaymath}
\Big[ \frac{\partial }{\partial t} + p(x)\Big] \sigma^1_{xx} = c_{11} \; \frac{\partial v_x}{\partial x},
\qquad
\Big[ \frac{\partial }{\partial t} + p(z)\Big] \sigma^2_{xx} = c_{13} \; \frac{\partial v_z}{\partial z},
\end{displaymath}
and the new equations for the $\sigma^1_{zz}$ and
$\sigma^2_{zz}$ decoupled stress tensors components equal to:
\begin{displaymath}
\Big[ \frac{\partial }{\partial t} + p(x)\Big] \sigma^1_{zz} =
c_{13} \; \frac{\partial v_x}{\partial x}, \qquad \Big[
\frac{\partial }{\partial t} + p(z)\Big] \sigma^2_{zz} = c_{33}
\; \frac{\partial v_z}{\partial z}.
\end{displaymath}
The decoupled shear $\sigma^1_{xz}$ and $\sigma^2_{xz}$ stress
tensor components follow the equations:
\begin{displaymath}
\Big[ \frac{\partial }{\partial t} + p(x)\Big] \sigma^1_{xz} = c_{44} \; \frac{\partial v_z}{\partial x},
\qquad
\Big[\frac{\partial }{\partial t} + p(z)\Big] \sigma^2_{xz} = c_{44} \; \frac{\partial v_x}{\partial z}.
\end{displaymath}
where now $\sigma_{xx} = \sigma^1_{xx} + \sigma^2_{xx}$,
$\sigma_{zz} = \sigma^1_{zz} + \sigma^2_{zz}$, and $\sigma_{xz}
= \sigma^1_{xz} + \sigma^2_{xz}$. These 5 linear \textbf{PDE}
equations will be used for a forward modeling in the following
section.

\section*{\textbf{Computational space–time implementation}}

\begin{figure}[htp]
\begin{center}
\includegraphics[width = 3.0 in, height= 3.0 in]{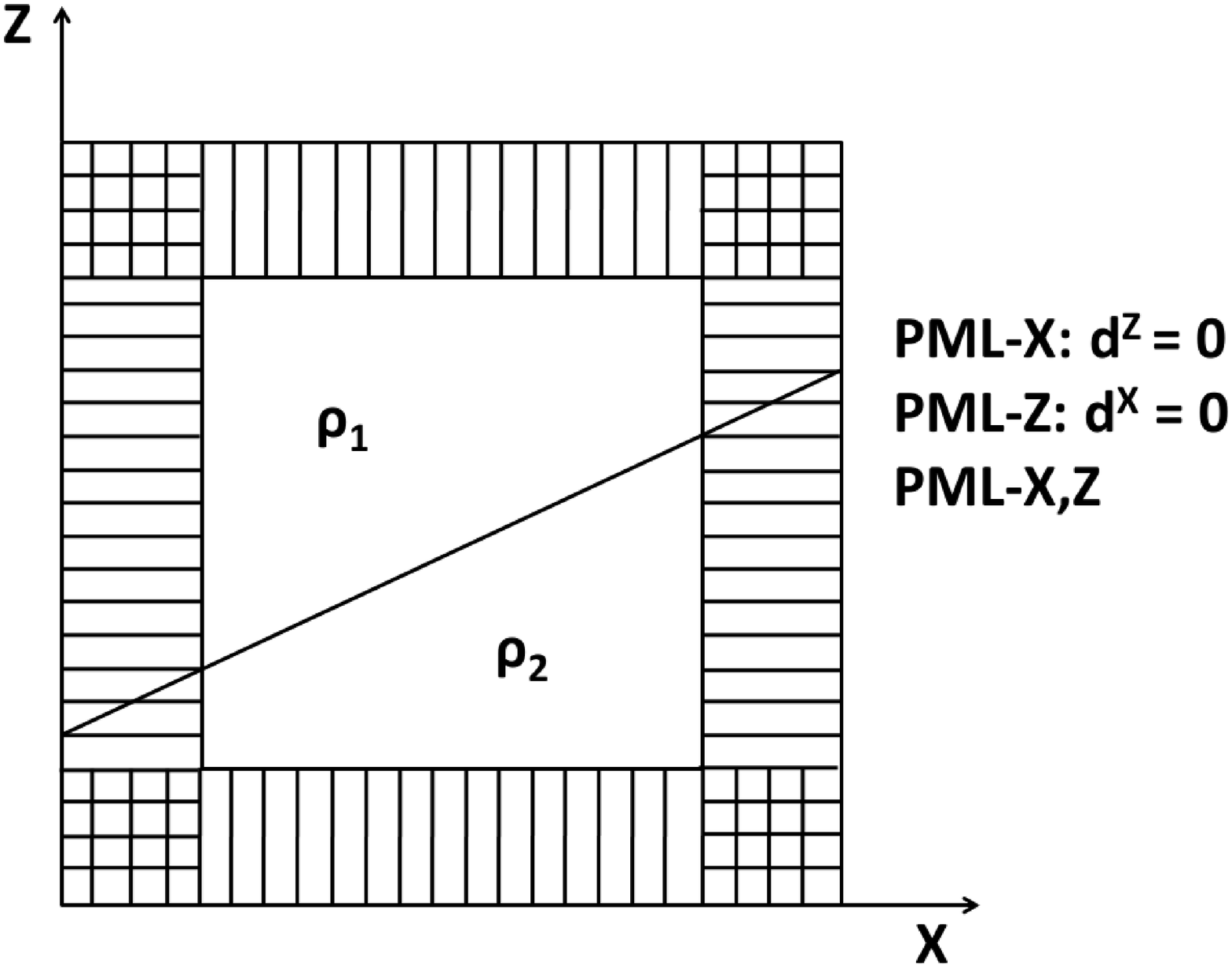}
\caption{\label{f0} Finite-difference simplified scheme showing the computational 2 two-layer dip model.
The vertical, horizontal and crossed lines near the borders show where the \textbf{PLM} are applied}
\end{center}
\end{figure}

\begin{figure}[htp]
\begin{center}
\includegraphics[width = 3.0 in, height= 3.0 in]{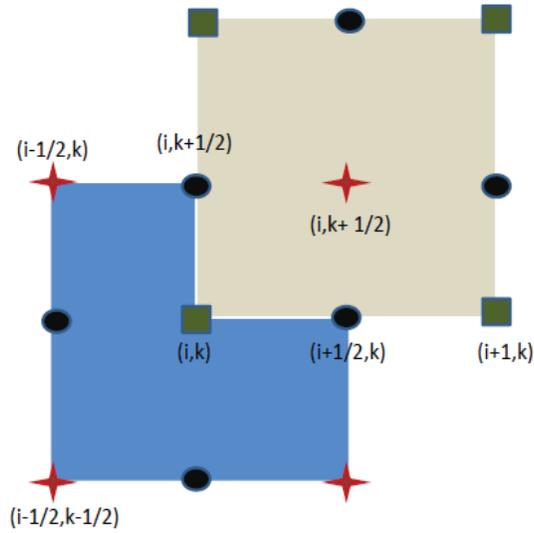}
\caption{\label{f1} Staggered finite-difference grid for
the velocity and stress updates, both stencils show the grids points where different fields components are calculated.
Velocity components are calculated in both stencils (black circles),
the shear stress components are calculated in the blue stencil (red starts), and the cinnamon stencil is used to calculate
the normal stress components (green squares).}
\end{center}
\end{figure}

The computational model that we propose in this simulation is
divided into a mesh of $N_x$ $\times$ $N_z$ points.
Fig.~(\ref{f0}) the finite-difference scheme shows the
computational edge domain where the \textbf{PML} are applied
according to the two-layer dip model. $\Delta x$ and $\Delta z$
are defined as the distances between the points in such a way
that $x = n_x \Delta_x$ y $z =n_z \Delta z$ with $n_x =
1...N_x$ y $n_z= 1...N_z$. For the step in time $\Delta t$ we
have that $t = n \Delta t$, being $n$ the step time (in
Fig.~\ref{f1}) the stencil time is not showed
\protect\cite{faria}.

The five variables $v^1_x$, $v^1_z$, $\sigma^1_{xx}$,
$\sigma^1_{zz}$, and $\sigma^1_{xz}$ are discretize into a two
dimensional staggered grid mesh, see Fig.~(\ref{f1}) for a
better explanation, where the velocity components are stored in
both stencils, the normal components of the stress tensors are
stored in one of the stencils, while the shear stress
components are located at another stencil. However, the major
difference of a staggered-grid scheme is that the velocity and
stress components are not known at the same grid point, as it
can be seen from Fig.~(\ref{f1}), henceforth we use a previous
scheme proposal with little modifications \protect\cite{faria}
where different stencils are used for normal and shear stress
field computations.

Following Fig.~(\ref{f1}) in the space domain $v_x$ is
calculated at the points $(i \pm 1/2, k)$, $v_z$ is calculated
at the points $(i,k \pm 1/2)$, the normal stress tensor
components $\sigma_{xx}$ and $\sigma_{zz}$ are calculated at
the points $(i,k)$ and finally the shear stress $\sigma_{xz}$
is calculated at the points $(i \pm 1/2, k \pm 1/2)$. The
density $\rho(x,y,z)$ is considered a constant density $\rho_0$
(see the last line of Table I for its correspondent values and
units) Physically it means that the \textbf{VTI} model is
homogeneous and that is does not take into consideration
earth-crust heterogeneities.

An explosive source is simulated using the amplitude of a
Ricker wavelet with a peak frequency $f_0 \sim 30$ Hertz.
Amplitudes are added to the velocities $v^ 1_x$ and $v^1_z$ at
each $n$ time-step. The stability limit of the discrete system
is given by the following condition:
\begin{center}
\[
\Delta t \; < 0.606 \; \frac{\Delta x}{V_p},
\]
\end{center}
where $V_p$ account s for the horizontal longitudinal speed
\protect\cite{faria}. The dispersion condition for the discrete
system is given by the inequality:
\begin{center}
\[
\Delta x \; < \; \frac{V_s}{10 \; f_0},
\]
\end{center}
where $V_s$ is the transverse wave speed. However some authors
\protect\cite{Becache} showed that the stability of the
classical \textbf{PML} model depends upon the physical
properties of the anisotropic medium and that it can be
intrinsically unstable. However the elastic constants in
table~\ref{t1} where previously proven\protect\cite{faria} to
meet Becache stability criteria. This algorithm could be
extended to 3-D situations by further studying the stability
and dispersion conditions.

Henceforward we proceed to obtain the discretization of the
velocity fields as follows for the decoupled finite-difference
system: the $v^1_x$ and the $v^2_x$ components are computed at
time-steps points $t^{n + 1}$ and $t^{n}$ (the stencil-time is
not shown here\protect\cite{faria}) with the following
finite-difference equations:
\begin{equation}
v^1_x |^{n+1}_{i+1/2,k} = v^1_x |^n_{i+1/2,k} e^{-p(x) \Delta t} + \frac{1}{\rho}\frac{\Delta t}{\Delta x}
e^{- 0.5 \; p(x) \; \Delta t} \Big[ \sigma_{xx}\Big|^{n+1/2}_{i+1,k} - \sigma_{xx}\Big|^{n+1/2}_{i,k} \Big],
\end{equation}
\begin{equation}
v^2_x |^{n+1}_{i+1/2,k} = v^1_x |^n_{i+1/2,k} e^{-p(z) \Delta t} + \frac{1}{\rho}\frac{\Delta t}{\Delta z}
e^{- 0.5 \; p(z) \; \Delta t} \Big[ \sigma_{xz}\Big|^{n+1/2}_{i+1/2,k+1/2} - \sigma_{xz}\Big|^{n+1/2}_{i+1/2,k-1/2} \Big].
\end{equation}

The $v^1_z$ and  $v^2_z$ components are calculated at the
times-step $t^{n + 1}$ and $t^{n}$ points with the help of
equations:
\begin{equation}
v^1_z |^{n+1}_{i,k+1/2} = v^1_z |^n_{i,k+1/2} e^{-p(x) \Delta t} + \frac{1}{\rho}\frac{\Delta t}{\Delta x}
e^{- 0.5 \; p(x) \; \Delta t} \Big[ \sigma_{xz}\Big|^{n+1/2}_{i+1/2,k+1/2} - \sigma_{xz}\Big|^{n+1/2}_{i-1/2,k+1/2} \Big],
\end{equation}
\begin{equation}
v^2_z |^{n+1}_{i,k+1/2} = v^1_x |^n_{i,k+1/2} e^{-p(z) \Delta t} + \frac{1}{\rho}\frac{\Delta t}{\Delta z}
e^{- 0.5 \; p(z) \; \Delta t} \Big[ \sigma_{xx}\Big|^{n+1/2}_{i,k+1} - \sigma_{xx}\Big|^{n+1/2}_{i,k} \Big].
\end{equation}

For the stress fields the discretization is as follows: First,
the decoupled equations for the normal stress tensor components
$\sigma^1_{xx}$ and $\sigma^2_{xx}$ are computed at the mesh
step-time points $t^{n \pm 1/2}$ using the equations:
\begin{equation}
\sigma^1_{xx}|^{n+1/2}_{i,k} = \sigma^1_{xx}|^{n-1/2}_{i,k} e^{-p(x) \Delta t} + C_{11} \; \frac{\Delta t}{\Delta x}
e^{- 0.5 \; p(x) \; \Delta t} \Big[ v_x \Big|^{n}_{i+1/2,k} - v_x\Big|^{n}_{i-1/2,k} \Big],
\end{equation}
and
\begin{equation}
\sigma^2_{xx}|^{n+1/2}_{i,k} = \sigma^2_{xx}|^{n-1/2}_{i,k} e^{-p(z) \Delta t} + C_{13} \; \frac{\Delta t}{\Delta z}
e^{- 0.5 \; p(z) \; \Delta t} \Big[ v_z \Big|^{n}_{i,k+1/2} - v_z \Big|^{n}_{i,k-1/2} \Big].
\end{equation}

Second, the decoupled equation for the normal stress components
$\sigma^1_{zz}$ and $\sigma^2_{zz}$ are computed at the time
points $t^{n \pm 1/2}$ by means of the following expressions:
\begin{equation}
\sigma^1_{zz}|^{n+1/2}_{i,k} = \sigma^1_{zz}|^{n-1/2}_{i,k} e^{-p(x) \Delta t} + C_{13} \; \frac{\Delta t}{\Delta x}
e^{- 0.5 \; p(x) \; \Delta t} \Big[ v_x\Big|^{n}_{i+1/2,k} - v_x\Big|^{n}_{i-1/2,k} \Big],
\end{equation}
and
\begin{equation}
\sigma^2_{zz}|^{n+1/2}_{i,k} = \sigma^2_{zz}|^{n-1/2}_{i,k} e^{-p(z) \Delta t} + C_{33} \; \frac{\Delta t}{\Delta z}
e^{- 0.5 \; p(z) \; \Delta t} \Big[ v_z\Big|^{n}_{i,k+1/2} - v_z\Big|^{n}_{i,k-1/2} \Big].
\end{equation}
Third, the decoupled equations for the shear $\sigma^1_{xz}$
and $\sigma^2_{xz}$ components are computed at time points
$t^{n \pm 1/2}$ by means of:
\begin{equation}
\sigma^1_{xz}|^{n+1/2}_{i+1/2,k+1/2} = \sigma^1_{zz}|^{n-1/2}_{i+1/2,k+1/2} e^{-p(x) \Delta t} + C_{44} \; \frac{\Delta t}{\Delta x}
e^{- 0.5 \; p(x) \; \Delta t} \Big[ v_z\Big|^{n}_{i+1/2,k} - v_z\Big|^{n}_{i-1/2,k} \Big],
\end{equation}
and
\begin{equation}
\sigma^2_{xz}|^{n+1/2}_{i+1/2,k+1/2} = \sigma^1_{zz}|^{n-1/2}_{i+1/2,k+1/2} e^{-p(z) \Delta t} + C_{44} \; \frac{\Delta t}{\Delta z}
e^{- 0.5 \; p(z) \; \Delta t} \Big[ v_x\Big|^{n}_{i,k+1/2} - v_x\Big|^{n}_{i,k+1/2} \Big].
\end{equation}

The elastic constants used for the two-layer model are given
according to table~\ref{t1}. These values correspond to
different types of \textbf{VTI} media and they are able to
accomplish the stability condition using a \textbf{PML}
computer domain \protect\cite{faria,dimitri,Becache}.

\begin{table}[hbt!]
\begin{center}\label{t1}
\caption{\textbf{VTI} elastic constants for model Fig.
\textbf{4(A)} \protect\cite{dimitri}}
\begin{tabular}{ccc}
\hline \hline
elastic constants ($\times 10^{10} N \; m^{-2}$) & upper layer & bottom layer \\ \hline \hline
$C_{11}$     & 16.5  & 16.7  \\ \hline
$C_{13}$    & 5.0  & 6.6 \\ \hline
$C_{33}$         & 6.2  & 14.0  \\ \hline
$C_{44}$       & 3.4  & 6.63  \\ \hline \hline
$\rho_0$ $(Kg \; m^{-3})$      & 7.100  & 3.200 \\ \hline \hline
\end{tabular}
\end{center}
\end{table}

\section*{\textbf{Analysis and Results}}
The present paper describes a new space-time 2-D methodology
with time-step and space-step control to solve 2-D \textbf{VTI}
linear decoupled \textbf{EWs} with \textbf{PML} conditions
using staggered-grids Fig.~\ref{f0} and Fig.~\ref{f1}.
Henceforth, we develop a computational model for an elastic
wave propagation simulation for 2-D \textbf{VTI} anisotropic
media, by combining a \textbf{FDTD} method on staggered grids
with a \textbf{PML} boundary condition. For that, a
High-Performance Linux application in \textbf{C} (over 1,000
lines of code) using Make, GDB and Valgrind's memcheck, to
generate and visualize 2-D response-impulses and synthetic
seismograms was developed.

\begin{figure}[ht]
\begin{center}
\includegraphics[width = 6.0 in, height= 8.0 in]{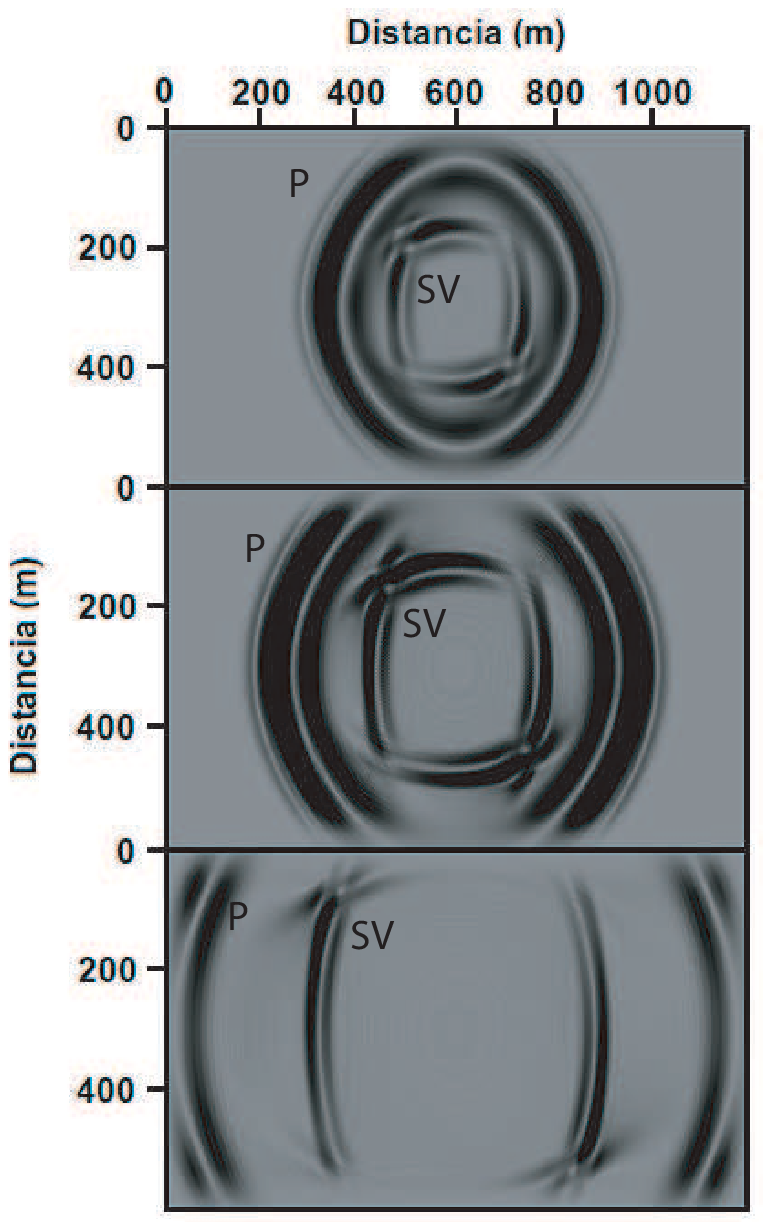}
\caption{\label{f2} Screen-shots at 0.2, 0.3 and 0.5 sec. are displayed for the vertical velocity wave-field component
propagating through a homogeneous medium with elastic constants from the upper layer given in Table I.
The triplication of the \textbf{SV} wave around the intermediate angles are observe to be asymmetrical for the \textbf{SV}
impulse-response. The absorbing effect of the \textbf{PML} at the edges of the computational domain is notable.}
\end{center}
\end{figure}

We establish that visualization of the impulse-response of the
\textbf{P} and \textbf{SV} modes are improved by the
\textbf{PML}. As a consequence, unwanted reflections from the
borders in Fig.~\ref{f2} are totally eliminated according to
Virieux \protect\cite{virieux}.

In Fig.~\ref{f2} simulations were performed with an explosive
source (Ricket-wavelength) centered in the middle of the model
to eliminate reflections from the edges. In Fig.~\ref{f2} time
screen-shots for $t = 0.2, 0.3$, and $0.5$ sec. are presented.
From those snapshots, we conclude that the \textbf{PML}
boundary conditions absorb the reflections from the border,
implying an exceptional numerical performance of the
computational \textbf{PML} technique. Another conclusion is
that the \textbf{P} 2-D response-impulse behave as a simpler
quasi-elliptical way in Fig.~\ref{f2}, while the \textbf{SV}
response-impulse presents triplications. Moreover, the effects
of the \textbf{VTI} elastic constants on \textbf{SV}
response-impulses affects the direction of propagation. We
observe from the snapshots in Fig.~\ref{f2} how the value of
the elastic constant $C_{13}$ affects the direction of the
\textbf{SV}. Triplications in the \textbf{SV} wavefront are
modeled. However, one question is not solve yet in this work,
the asymmetric behavior of the \textbf{SV} impulse-response in
Fig.~\ref{f2} with respect to the vertical axis.

\begin{figure}[ht]
\begin{center}
\includegraphics[width = 7.0 in, height= 7.0 in]{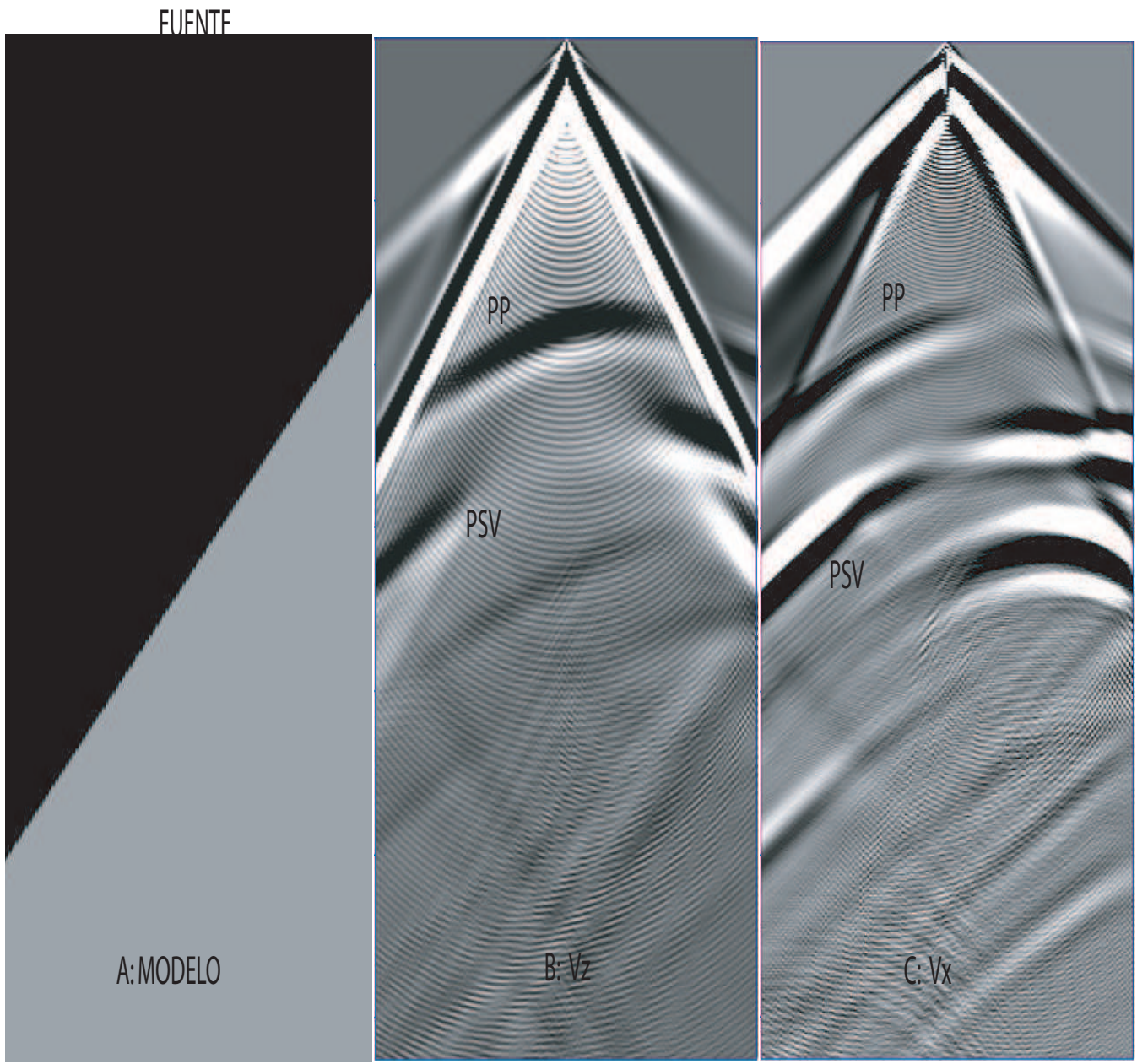}
\caption{\label{f3} Two-layers \textbf{VTI} dip model \textbf{4A}. Synthetic seismograms
for the vertical \textbf{Vz} are presented in \textbf{4B}. Synthetic seismograms for the horizontal \textbf{Vx} are displayed in \textbf{4C}.
In \textbf{4B} and \textbf{4C} screen-shops for the \textbf{PP} and the \textbf{PSV} reflection events due to the dip are presented.}
\end{center}
\end{figure}

Subsequently, the explosive source is relocated to the upper
central part of the 2-D model sketched in Fig.~\ref{f3}
\textbf{A}. On the surface, fifty receivers are uniformly
distributed on both sides of the source. We see how the
velocity fields cancel out at the edges of the model in
Fig.~\ref{f3} \textbf{A}. Then we obtain the synthetic
seismograms for the two layers model, each of them has
different elastic constants values. The synthetic seismograms
of the vertical component $V_z$ (Fig.~\ref{f2} \textbf{B}) and
the horizontal component $V_x$ (Fig.~\ref{f2} \textbf{C}) are
shown with the primary reflections events due to the dip layer.

Henceforth, it remarkably shows the effectiveness of the
\textbf{PML} using the decoupled linear system of equations
(1)-(10) for 2-D \textbf{VTI} media simulation in Oil and Gas
R\&D.

\section*{Acknowledgments}
One of the authors, P. Contreras wishes to express his
gratitude to Dr. V. Grechka for pointing out an invariant
symmetric question for the \textbf{SV} impulse-response. The
authors also acknowledge Drs. E. Sanchez, and Prof. J. Moreno
for several discussions regarding the numerical implementation
of this work.

\newpage

\end{document}